\documentclass[11pt]{article} 
\pdfoutput=1
\usepackage[margin=1in]{geometry}
\usepackage{authblk}
\usepackage{amsmath,amsfonts,amssymb,amsthm,hyperref,color,graphics,graphicx,subfig,float}
\usepackage{enumerate}
\usepackage{newlfont}
\usepackage{bibentry}
\usepackage{algorithm}
\usepackage{algorithmic}
\usepackage{graphicx}
\usepackage[utf8]{inputenc}
\usepackage{balance}
\usepackage{placeins}
\usepackage{pbox}

\begin{document}
\title{Modeling and Performance Comparison of  Privacy Approaches for Location Based Services}
 \author[1]{Pratima Biswas}
 \author[2]{Ashok Singh Sairam}
\affil[1]{Indian Institute of Technology Patna, pratima.pcs13@iitp.ac.in}
\affil[2]{Indian Institute of Technology Guwahati, ashok@iitg.ernet.in}
\maketitle

\begin{abstract}
In pervasive computing environment, Location Based Services (LBSs) are getting popularity among users because of their usefulness in day-to-day life. LBSs are information services that use geospatial data of mobile device and smart phone users to provide information, entertainment and security in real time. A key concern in such pervasive computing environment is the need to reveal the user's exact location which may allow an adversary to infer private information about the user. To address the privacy concerns of LBS users, a large number of  security approaches have been proposed based on the concept of \textit{k}-anonymity. The central idea in location \textit{k}-anonymity is to find a set of \textit{k-1} users confined in a given geographical area of the actual user, such that the location of these \textit{k} users are indistinguishable from one another, thus protecting the identity of the user. Although a number of performance parameters like success rate, amount of privacy achieved are used to measure the performance of the \textit{k}-anonymity approaches, they make the implicit, unrealistic assumption that the \textit{k-1} users are readily available.  As such these approaches ignore the \textit{turnaround} time to process a user request, which is crucial for a real-time application like LBS. In this work, we model the \textit{k}-anonymity approaches using queuing theory to compute the average sojourn time of users and queue length of the system. To demonstrate that queuing theory can be used to model all \textit{k}-anonymity approaches, we consider  \textit{graph-based}  \textit{k}-anonymity approaches. The proposed analytical model is further validated with experimental results.

\end{abstract}

\section{Introduction}
With the wide availability of location-aware devices and advancement of positioning technologies like Global Positioning Systems (GPS) to determine exact locations of users and objects of interest, a new class of applications called location-based services (LBS) have become highly popular. These applications can vary from utility applications like finding points of interest, friends currently present in ones vicinity to serious applications like sending alarm messages during emergency etc \cite{survey1,survey2}. One of the main concern in using these services, is that it require revealing ones location which may allow an adversary to infer sensitive information of the user. To address the privacy concerns of LBS, a number of approaches have been proposed, popular among them are those approaches that implement the concept of \textit{k}-anonymity. The key idea is to find a set of $k$ users confined in a given geographical area such that they are indistinguishable from one another, thus protecting the identity of the user.  

Central to the idea of \textit{k}-anonymity in LBS is a trusted third party (TTP) which is delegated with the task of anonymization.  When a LBS query arrives at the TTP, it finds $k-1$ other users in the vicinity of the user and sends the obfuscation area to the LBS server. This is known as \textit{cloaking} or position \textit{obfuscation}.  The different LBS privacy approaches using \textit{k}-anonymity basically differ in the way how the $k-1$ other users are selected. However, the underlying goal of all these approaches is to send the minimum possible \textit{cloaking} area to the LBS server. The top-down approach \cite{Quadivison, top-down2} and bottom-up approach \cite{Casper,bottom-up2} generate the cloaking area by inspecting the \textit{quadrants} which are in the vicinity of the user.  In both the approaches, the cloaking area returned is not guaranteed to be optimal.  In the graph-based approach \cite{CliqueCloak,iClique}, the cloaking area is generated by forming a graph among the users who have issued queries. For example in CliqueCloak\cite{CliqueCloak} approach, whenever a query arrives it is checked if the location point of the user forms a  clique with $k-1$ other users. This approach guarantees that the cloaking area formed is the minimum for a given set of users.  In this work, we are primarily concerned with \textit{graph-based} \textit{k}-anonymity approaches.

The \textit{k}-anonymity privacy approaches available in literature make an unrealistic, implicit assumption that the $k-1$ other users are readily available.  However, in practice queries for anonymization will arrive at unpredictable times and when they arrive other users may not be available. In such a case, the first $k-1$ users will always have to wait. Thus the natural question that arise is how long a LBS query may have to wait  before it can be serviced, for how long will the TTP be busy in computation and so on. The existing  \textit{k}-anonymity approaches consider different performance parameters like success rate, amount of privacy level achieved, etc. but do not consider parameters like  average response time of a query which is very important as the queries are fired in real time and users want fast response.

The contribution of our work can be summarized as follows:
\begin{enumerate}[i]
\item{{Modeling of \textit{k}-anonymity privacy approach:} Our first aim is to develop a mathematical framework that can be used to evaluate graph-based \textit{k}-anonymity privacy approaches in terms of request-to-response time of a query, the number of queries present in the TTP, length of a busy period and length of an idle period.}
\item{{Experimental validation of the mathematical model:} The second goal of our work is to experimentally compute the performance parameters estimated using our mathematical model and compare the results.}
\item{{Comparison:} The last objective is to compare the various graph-based $k$-anonymity approaches available in literature based on our proposed model.}
\end{enumerate}

In this work, we use the concept of single service systems and bulk service systems of \textit{Queueing theory} to model the  \textit{k}-anonymity LBS privacy approaches that uses a TTP.  The top-down and bottom-up \textit{k}-anonymity privacy approaches have a non-deterministic processing time, thus they can be modeled by single service systems. In the graph-based approach, once a graph is successfully formed all the $k$ queries involved in the graph are simultaneously processed, thus this approach can be modeled using the bulk service systems. However, the presence of $k$ queries does not necessarily ensure that all the queries will be successfully anonymized since they may not form the desired graph. Thus we need to design a variant of the bulk query processing model by incorporating an anonymizing probability. Results show that our mathematical model as compared to experimental results has a high accuracy with an error percentage of about 2.5 percent. \\


The remainder of our work is organized as follows. The background of our work is presented in Section \ref{background}. In section \ref{queue} we describe how queuing theory can be used to model the \textit{k}-anonymity privacy approaches. The proposed queuing model  is given in section  \ref{proposed_model}.   Experimental results are given in section \ref{expts} and finally the concluding remarks are given in section \ref{conclusion}.

\section{Background}\label{background}
LBSs \cite{C1} are information services that exploit a mobile users current location to provide value added information. The  basic components of a LBS system are \textit{mobile devices}, \textit{anonymizer} or \textit{trusted third party} (\textit{TTP}), \textit{LBS server} or \textit{service provider} and the \textit{content provider} as shown in Figure \ref{LBS_model}. This model matches most approaches described in literature. The mobile devices are tools used by users to access LBS services, to send requests and retrieve results. Such devices can be Personal Data Assistants (PDAs), laptops or mobile phones. The user sends the location-based query to the LBS server through a  communication network. The query is in  the form of $(id, l, q)$, where $id$ is the identity of the user, $l$ is  the user's current location and $q$ represents the query content.  The user position is determined by using positioning technology such as the Global Positioning System (GPS). Service provider maintain various services to offer different kinds of LBS services to users. They are responsible for processing service requests and sending back query results to the mobile users. The service providers calculate positions, search for a route, or search specific information based on the user’s position. Service providers usually do not store information, instead content providers are responsible for collecting and storing geographic data, location-based information, and other related data. These data are requested and processed by the service providers and then returned to users.

\subsection{Privacy Mechanism for LBS}
\begin{figure}[h!]
 \begin{center}
 \includegraphics[scale=0.4]{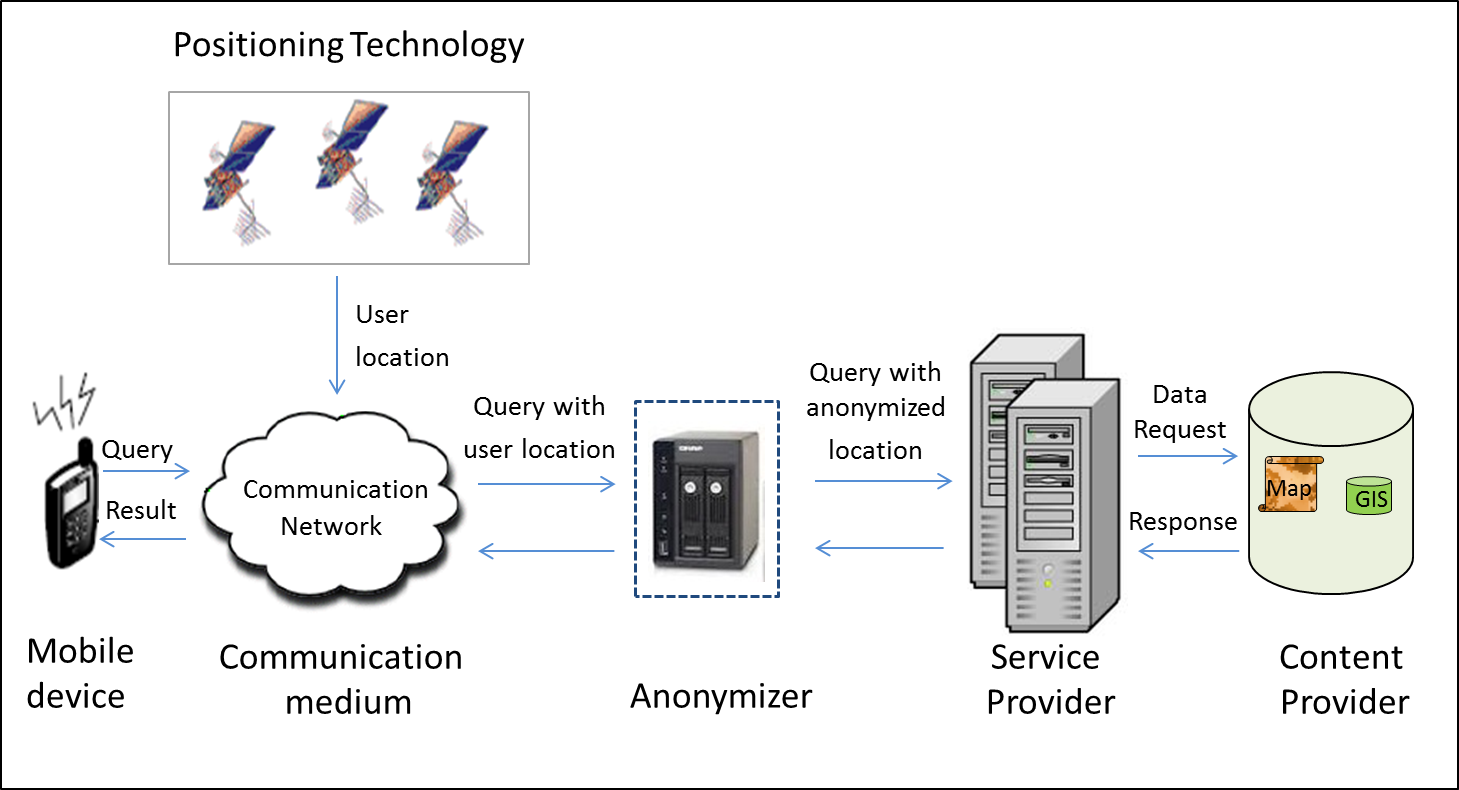}	
 \caption {A generic location-based service model}
\label{LBS_model}
\end {center}
\end {figure}

Although LBS provide useful personalized service for mobile users, but these services raise a serious privacy concern of the users, as they need to reveal their location information to the LBS server. A common technique to provide location privacy to the mobile user is to introduce a trusted third party (TTP) server known as \textit{Anonymizing Server} between mobile device and LBS server as shown in Figure \ref{LBS_model}. The two most commonly used privacy metric by the TTP are \textit{cloaking} and \textit{location k}-anonymity. In cloaking, the location information of the user is protected by providing a lower resolution in terms of space and time. That is instead of providing the exact location a larger region and temporal information is reported. In location \textit{k}-anonymity, the TTP makes the location information of the user indistinguishable from that of at least $k - 1$ users.  

 The LBS query after incorporating these privacy mechanisms is formally represented as \cite{iClique}:
\begin{equation}
 q:[(u_{id},q_{no}),(x,y,t),k,(d_x,d_y,d_t),C]
\label{lbs_query}
 \end{equation}

where $u_{id}$ and $q_{no}$ are user identification number and query reference number respectively. This pair combinedly identify a LBS query uniquely.  The coordinate $(x,y)$ indicate the current location of user at time stamp $t$. The parameter  $k$ is the minimum desired anonymity level. The variables  $d_x$ and $d_y$ indicate the spatial tolerance in the query result and $d_t$ specifies the temporal tolerance. $C$ refers to the content of the query issued by the client.

\section{Modeling LBS Privacy Approaches using M/G/1 Queueing System } \label{queue}
\begin{figure}[h!]
\begin{center}
\includegraphics[scale=0.56]{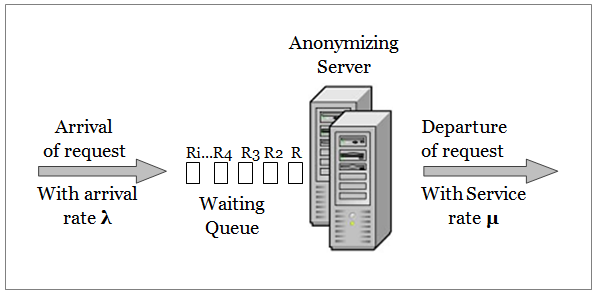}	
 \caption {System Model}
 \label{System Model}
 \end {center}
\end {figure}

In this work, we model LBS privacy approaches using \textit{Queueing theory}, so as to analyze the algorithm of these approaches and compare their performance by evaluating various performance metrics.  The generic LBS service model shown in Figure \ref{LBS_model} can be represented by a more specific system model shown in Figure \ref{System Model}. In the system model, the TTP is replaced by a \textit{queueing system}. In the proposed system model, LBS requests issued by mobile users arrive into the  queueing system according to a $Poisson$ process with mean arrival rate $\lambda$ customers per unit time and wait in a queue for being anonymized. The customers in the queue are served in a  First Come First Served (FCFS) manner with mean service rate $\mu$. 

The service rate of the LBS server  is assumed to be constant. However, the service time for each customer is additionally determined by an external environment, for instance the probability of occurrence of $k-1$ other users. Therefore, the queries require random amount of time to complete. The service time depends on the particular privacy approach being used. Different privacy approaches need different time to anonymize a query, with each approach having standard deviation $\sigma$ from its mean processing rate $\mu$. Thus the service time for requests is described by the General distribution. Each arriving query brings a certain amount  of work having general distribution with mean $E(\tilde{x})$. The M/G/1 queueing model can be best used to describe the $k$-anonymity privacy system where arrivals are Markovian, service times have a General distribution and there is a single server. The TTP employed in LBS system acts as a single queueing server where the inter-arrival times of the LBS requests are independent random variables and the occurrence of arrival  is according to a Poisson process. The maximum queue length and population size of the system can theoretically raise up-to infinity.

In order to conform to queueing theory, we assume that the service rate is always greater than the arrival rate. We follow this assumption both in our theoretical as well as practical model.


\subsection{$k$-Anonymity}
In the previous section, we had seen that the arrival of queries in a LBS system is non-deterministic and it can in general be assumed that the arrivals follow a Poisson process. The difference between \textit{k}-anonymity privacy approaches is the manner in which the query is processed. Thus the  \textit{k}-anonymity approaches can be characterized by their processing time. Quad-tree based privacy can be broadly classified as top-down \cite{Quadivison, top-down2} and bottom-up \cite{Casper,bottom-up2}.  In a top-down approach like Quad-division\cite{Quadivison}, the area around the user that issued the query is subdivided into quadrants until the number of users fall below $k$. The previous quadrant that fulfills the privacy requirement is then returned. Casper\cite{Casper} is a representative of the bottom-up approach. The algorithm checks if the number of users in the initial area (called cell) satisfies the privacy requirement, if so the cell is returned. Otherwise the adjacent cells are considered and the approach is continued recursively until number of users in the combined cell becomes greater than or equal to $k$. In both the approaches, it is assumed that the number of registered users is greater than $k$.  These approaches have a non-deterministic processing time and thus can be modeled by a M/M/1 queueing model \cite{book}.  

B. Bamba et. al \cite{PrivacyGrid} argue that depending on the scenario, either top-down or bottom-up approach will perform anonymization faster. Thus they propose a hybrid approach that combines the strength of both the approaches, in order to further reduce the anonymization time. The input to the algorithm is the required anonymity level $k$ and spatial  diversity $l$. The authors argue that for lower value of $k$ and higher spatial resolution value, a bottom-up approach is beneficial. On the other hand, for higher $k$ and lower spatial resolution, a top-down approach works faster. Thus such  hybrid approaches can also be modeled using the M/M/1 queueing model.

The LBS privacy schemes that we have seen so far assume that the adversary can acquire a single request, the so called \textit{snapshot} case. Historical attacks \cite{Historical,Historical2, Hilbert,Hilbert2} assume that the adversary is able to link a set of requests fired by a user.  One solution to this problem is the application of \textit{Hilbert Cloaking} algorithm \cite{Hilbert, Hilbert2} which basically exploit the Hilbert space filling curve to find a total order among the user's various locations. The algorithm then returns the minimum bounding rectangle (MBR), considering the position of other users that are in the same MBR as the actual user. To improve the spatial accuracy, heuristics have been proposed   \cite{Historical,Historical2} that recursively tries to shrink the obtained MBR. From a queueing theory point of view, these algorithms are similar to the top-down approach and thus can be modeled using a  M/M/1 queueing model. 

A user's location  carries much more information then just the coordinate values. The \textit{semantic} of a location defines the criticality of the position information such as  hospitals, restaurants, etc.  A user may not mind sharing his location information as long as he does not enter such semantic zones . Map-aware obfuscation algorithms \cite{map-aware} divide the area into cells, and if the user's position is in a semantically sensitive cell, it obfuscates the cell by adding additional cells. Thus the basic principle of these class of algorithms are similar to that of the top-down approach. Although these class of algorithms do not fall in the category of spatial $k$-anonymity, they are nevertheless complementary.

A major challenge in designing cloaking algorithms is to find the smallest cloaking area for a given spatial and temporal tolerance. The CliqueCloak \cite{CliqueCloak} approach address these issues by considering the queries from different clients and forming edges between them, if the corresponding clients are within each others range. It checks if a k-clique is formed, compute the minimum bounding rectangle of the messages within the clique and return it as the cloaking area. From a queueing theory point of view, the TTP can process the request only if it receives $k$ queries. However, the presence of $k$ queries itself does not ensure that processing will be successful since the queries may not form a k-clique. 

X. Pan et. al.\cite{iClique}  show that the privacy aware approaches consider only the present location of a user but does not consider its mobility. The authors show that if an user frequently fires queries, the location of the user can be inferred to be within a region, a so called maximum movement boundary (MMB). To counter such location dependent attacks, they propose a cloaking algorithm iClique \cite{iClique}. The approach is similar to the CliqueCloak as the algorithm also checks if a k-clique is formed. However, the edges between nodes are formed if and only if they are within the MMB of each other. Thus we can also model iClique approach using a variant of the bulk queue processing model.

\section{Queueing model for Existing $k$-Anonymity Privacy Approaches} \label{proposed_model}
The standard M/G/1 Queueing model can be used to model most of the existing privacy approaches based on $k$-anonymity mechanism to provide user privacy in location based services.  The M/G/1 queueing system is a single-server queueing system with Poisson input, general service time distribution and unlimited number of waiting positions. 

In this section, we model the ClickCloak as they cannot be modeled by any of the standard bulk processing model. The model for ClickCloak can be easily extended to model iClique.

The $traffic$ $intensity$ $\rho$ is defined as the ratio of the mean service time to the mean inter-arrival time of the customer, that is $\rho=\lambda/\mu$. For stability of the system, the condition $\rho$ $<$ 1 must be satisfied.

\subsection{Queueing model for \textit{CliqueCloak} Privacy Approach}
In the \textit{CliqueCloak} approach\cite{CliqueCloak}, incoming queries into the TTP are represented as vertices of a  graph. An edge is formed  between two vertices if they are in the spatial tolerance region of one another.  The authors prove that  for any incoming query, a set of $M$ queries will be its valid $k$-anonymous perturbation if and only if the set forms a clique of size $k$. Thus the main working principle of this approach is to check whether the incoming queries form a clique of size $k$. The procedure of finding cloaking set of size $k$ is shown in algorithm {\ref{clique_algo}}. In the algorithm, $d_{qn}$ and $d_{avg}$ denote the distance between node $q$ and $n$ and the average distance respectively.
\vspace{0.2in}
\begin{algorithm}
 \caption{Clique-Cloak}
 \label{clique_algo}
 \begin{algorithmic}
 \STATE Initialize quadrant $A$ as Total area covered by anonymizer
\STATE Generate request $q$ with location $(x,y)$
\STATE $G$: Constraint graph
\STATE Add $q$ as node to graph $G$

\IF{$|q|$ $\geq$ k} 
 \FOR{\textbf{each} node $n \in {G}$}
   \IF{$d_{qn} \leq d_{avg}$} 
 \STATE(add edge between $q$ and $n$) 
 \ENDIF
 \ENDFOR
 \IF{clique of size $k$ exist}
 \STATE return cloak set $C$ of size $k$
\ENDIF 
\ENDIF  

\end{algorithmic}
\end{algorithm} 

The CliqueCloak system is a bulk processing approach as it returns a cloak set composed of $k$ queries. However, this approach cannot be modeled by the bulk processing queueing model, $M/M^k/1$ has the presence of $k$ queries, but it does not necessarily mean that all $k$ of them will  be successfully processed. The queries must further satisfy a constraint  for successful anonymization and the processed bulk of requests will leave the system .Otherwise it is considered as unsuccessful attempt to anonymize the bulk of requests. The requests will remain in the system to get anonymized with new arrivals. Thus to model the  CliqueCloak approach, we need to develop a new queueing model that considers the probability distribution of the service request.  On the lines of queueing theory,  we follow the following steps for modeling the CliqueCloak privacy system
\begin{itemize}
\item Construction of Markov chain model for CliqueCloak system
\item Formation of system state equation for Markov model 
\item Determination of probability distribution function by solving the equations
\item Derivation of expression for the performance matrices.
\end{itemize}

In the sections below, we explain each of these steps in detail.\\

\subsubsection{\bf {Construction of Markov Chain}}
\begin{figure}[h!]
\begin{center}
\includegraphics[scale=.36]{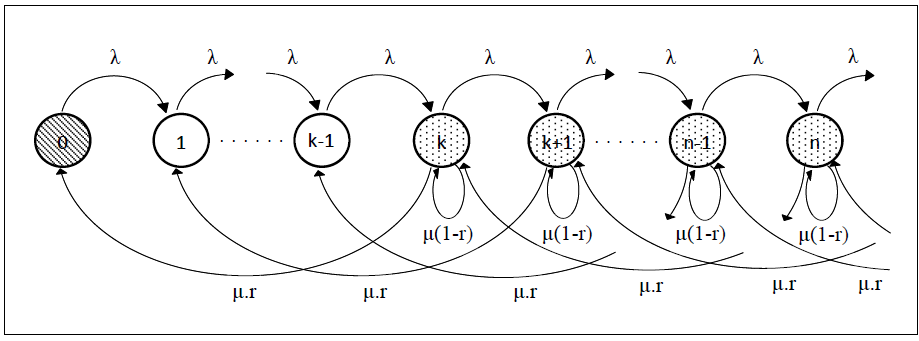}	
 \caption {State-transition-rate diagram}
 \label{diagram}
 \end {center}
\end {figure}

To model the privacy system, the very first task is to construct a Markov chain which depicts the state transition behavior of the CliqueCloak privacy system. The state-transition-rate diagram  is shown in Figure {\ref{diagram}} in which the states are labeled with a positive integer $n$.  For each arrival of request with arrival rate  $\lambda$ there is a transition of system from state $n$ to state $n+1$ as the number of requests is increased by 1. 


Initially the system is in state 0 as there is no request in the system. With the arrival of first request, the system state changes from 0 to 1 and the query waits in a queue for service till the occurrence of $k$ arrivals, where $k$ is the minimum bulk size to be served together.  After the arrival of $k$ requests, all the requests are processed together with a processing rate $\mu$ to find a clique of size $k$. If a clique is found, the anonymization is successful and all the $k$ requests leave the system. Hence the state of the system changes to $n-k$. On the other hand if a clique is not formed it leads to unsuccessful attempt for anonymizing the requests. The system waits for the next arrival of request. The state of the system remains unchanged.

In Figure {\ref{diagram}}, if $r$ denotes probability of being anonymized  then the term $\mu .r$ with backward arrow indicates that if the bulk of requests is processed and anonymized successfully then the system moves back by $k$ state. However the loop labeled by $\mu (1-r)$ specifies that if the requests are processed but does not get anonymized, then the system remains in same state.

{\bf Computation of Anonymizing Probability $r$:}
The proposed model based is the generic framework for all clique based privacy approach with difference in their anonymizing probability $r$.
The next obvious question that arises is how to determine the anonymization probability $r$ different approaches.
For an illustration we will determine $r$ for CliqueCloak and 
iClique privacy approaches. Consider any two nodes $p$ and
$q$ with locations $(x_p,y_p)$ and $(x_q,y_q)$ respectively in the area of size $X\text{ x }Y$ . Similarly let $(dx_p,dy_p)$ and $(dx_q,dy_q)$ be the spatial 
tolerance of queries  $p$ and $q$ as shown in equation \ref{lbs_query}.  In the CliqueCloak approach an edge will be formed 
between $p$ and $q$ if the coordinate $(x_p,y_p)$ lies in the area $dx_q \text{ x } dy_q$ and the coordinate $(x_q,y_q)$ lies 
in the area $dx_p \text{ x } dy_p$. The probability of the coordinate $(x_p,y_p)$ lies in the area 
$dx_q \text{ x } dy_q$ over the area $X\text{ x }Y$ is $(dx_q \text{ x } dy_q)/X\text{ x }Y$. similarly for coordinate $(x_q,y_q)$ 
the probability will be $X\text{ x }Y$ is $(dx_p \text{ x } dy_p)/X\text{ x }Y$.
Thus the probability of an edge being formed between nodes $p$ and $q$ is

\begin{equation*} \label{}
prob(p,q)=\frac{(dx_q \text{ x } dy_q)(dx_p \text{ x } dy_p)}{{(X \text{ x } Y)}^2}
\end{equation*}

where the region $X \text{ x } Y$ is the total area under consideration. Hence the probability of having clique of size $k$ is determined as:

\begin{equation*} \label{}
r_{CliqueCloak}=\frac{1}{{(X \text{ x } Y)}^{2k}}\prod_{i=1}^k(dx_i \text{ x } dy_i)^{k-1}
\end{equation*}

In the iClique \cite{iClique} privacy approach, an edge is formed between any two nodes $p$ and $q$, if $p$ lies in the maximum movement boundary (MMB) of $q$ and vice versa.Therefore the anonymizing probability $r$ for iClique is given as:

\begin{equation*} \label{}
r_{IClique}=\frac{1}{{(X \text{ x } Y)}^{2k}}\prod_{i=1}^k(MMB_i)^{k-1}
\end{equation*}

\subsubsection{\bf Formation of System State Equation}
Let $P_n$ represents probability of having $n$ requests in the system. In other words $P_n$ is the probability of being in the state $n$. Now we will write down system equation for each state describing the motion of the system  using  {\it flow conservation law} [10]. The flow conservation law states that - in equilibrium the probabilistic flow rate into a state must  be equal to the probabilistic flow rate out of that state. For instance consider state 0, the flow rate into the state 0 (indicated by  arrow entering into the state 0 in Figure{\ref{diagram}}) is $P_0{\lambda}$ and the flow rate out of the state 0 (indicated by the arrow  leaving the state 0) is $P_k{\mu}r$. Therefore according to the flow conservation law, 
\begin{equation}
 P_n{\lambda}=P_k{\mu}r {\label{a1}}, \hspace{.450in} n=0 \hspace{1.1in}
\end{equation}
Observe that the incoming and the outgoing patterns for state $1$ to $k-1$ are identical. Therefore the system state equation for state $1$ to $k-1$ 
are same and obtained as,  
\begin{equation}
P_n{\lambda}=P_{n-1}{\lambda}+P_{n+k}{\mu}r {\label{a2}}, \hspace{.45in} 0<n<k \hspace{.3in}
\end{equation}
Similarly the equation for remaining states is given by,
\begin{equation}
P_n({\mu}r+{\lambda})=P_{n-1}{\lambda}+P_{n+k}{\mu}r {\label{a3}} \hspace{.45in} n{\geqslant}k\hspace{.3in}
\end{equation}
The equations {\ref{a1}}, {\ref{a2}} and {\ref{a3}} are the obtained system state equations for the mentioned state-transition-rate diagram of CliqueCloak privacy system.

\subsubsection{\bf {Determination of probability distribution function $P_n$}}
Now we will solve the system of equations to determine $P_n$, the probability distribution for number of requests in the system using 
{\it z-Transformation} method which involves following two steps

\begin{itemize}
\item Applying z-Transformation to determine P(z)  
\item Applying inverse z-Transformation to determine $P_n$
\end{itemize}
{\bf {Determination of P(z) : }}
Equations {\ref{a2}} and {\ref{a3}} are almost identical except for the term  $\mu rP_n$ in the latter; consequently let us operate z-transform upon these equations in the range $n$
\begin{equation*}
\begin{split}
\sum_{n=1}^{\infty}P_n({\mu}r+{\lambda})z^n-\sum_{n=1}^{k-1}P_n{\mu}rz^n=\sum_{n=1}^{\infty}P_{n-1}{\lambda}z^n+\\
\sum_{n=1}^{\infty}P_{n+k}{\mu}rz^n 
\end{split}
\end{equation*}
Rearranging the above equation we will get
\begin{equation*}
\begin{split}
({\mu}r+{\lambda})\sum_{n=1}^{\infty}P_nz^n-{\mu}r\sum_{n=1}^{k-1}P_nz^n={\lambda}z\sum_{n=1}^{\infty}P_{n-1}z^{n-1}+\\
\frac{{\mu}r}{z^k}\sum_{n=1}^{\infty}P_{n+k}z^{n+k}
\end{split}
\end{equation*}
Let  $P(z)=\sum_{n=0}^{\infty}P_{n}z^n$, substituting this value in the above equation we get
\begin{equation*}
\begin{split}
({\mu}r+{\lambda})[P(z)-P_0)]-{\mu}r\sum_{n=1}^{k-1}P_nz^n={\lambda}zP(z)+\\
\frac{{\mu}r}{z^k}[P(z)-\sum_{n=0}^{k}P_{n}z^{n}]
\end{split}
\end{equation*}
\begin{equation*}
\begin{split}
[{\mu}r+{\lambda}-{\lambda}z-\frac{{\mu}r}{z^k}]P(z)=({\mu}r+{\lambda})P_0+{\mu}r\sum_{n=1}^{k-1}P_nz^n-\\
\frac{{\mu}r}{z^k}\sum_{n=0}^{k}P_{n}z^{n}
\end{split}
\end{equation*}
\begin{equation*}
\begin{split}
[{\mu}r+{\lambda}-{\lambda}z-\frac{{\mu}r}{z^k}]P(z)=({\mu}r+{\lambda})P_0+\\
{\mu}r\sum_{n=1}^{k-1}P_nz^n(1-\frac{1}{z^k})-\frac{{\mu}r}{z^k}P_{0}-{\mu}rP_k
\end{split}
\end{equation*}
Using equation {\ref{a1}}
\begin{equation*}
\begin{split}
[{\mu}r+{\lambda}-{\lambda}z-\frac{{\mu}r}{z^k}]P(z)=({\mu}r+{\lambda})P_0+\\{\mu}r\sum_{n=1}^{k-1}P_nz^n(1-\frac{1}{z^k})-\frac{{\mu}r}{z^k}P_{0}-P_0{\lambda} 
\end{split}
\end{equation*}
\begin{equation*}
[{\mu}r+{\lambda}-{\lambda}z-\frac{{\mu}r}{z^k}]P(z)={\mu}r\sum_{n=0}^{k-1}P_nz^n(1-\frac{1}{z^k})
\end{equation*}
\begin{equation} \label{c}
P(z)=\frac{r(1-z^k)\sum_{n=0}^{k-1}P_nz^n}{z^{k+1}\rho+r-(r+\rho)z^k}
\end{equation}
	
{\bf {Determination of ${P_n}$ : }}Now we will apply inverse z-transform on $P(z)$ to obtain the required expression  
for probability distribution function of number of requests in the system.
\begin{equation} \label{d}
\frac{z^{k+1}\rho+r-(r+\rho)z^k}{(1-z)(1-z/z_0)}=C\sum_{n=0}^{k-1}P_nz^n
\end{equation}
where $C$ is a constant to be evaluated below. Using Equation {\ref{d}} we can rewrite the Equation {\ref{c}} as
\begin{equation} \label{e}
P(z)=\frac{r(1-z^k)}{C(1-z)(1-z/z_0)}
\end{equation}
Applying L'Hospital's[9] rule and putting $P(1)=1$ in equation {\ref{e}} we have
\begin{equation}
C=\frac{rk}{(1-1/z_0)}
\end{equation}
On putting the value of $C$ in equation {\ref{e}}, it becomes
\begin{equation} \label{f}
P(z)=\frac{(1-z^k)(1-1/z_0)}{k(1-z)(1-z/z_0)}
\end{equation}
 now we carry out the partial fraction expansion on equation {\ref{f}} to apply inverse transform
\begin{equation} \label{}
P(z)=(1-z^k)[\frac{1/k}{(1-z)}-\frac{1/kz_0}{(1-z/z_0)}]
\end{equation}
\begin{equation*} \label{}
P_n=f_n-f_{n-k}
\end{equation*}
\begin{equation*} \label{}
f_n=\frac{1}{k}[1-z_0^{-n-1}]
\end{equation*}
Similarly,
\begin{equation*} \label{}
f_{n-k}=\frac{1}{k}[1-z_0^{-n+k-1}]
\end{equation*}
Therefore
\begin{equation} \label{g}
P_n=\frac{1}{k}(z_0^{k-n-1}-z_0^{-n-1})
\end{equation}
We discussed above that  $z_0$ is one of the root of the denominator of Equation {\ref{c}}. Thus the expression $z^{k+1}\rho+r-(r+\rho)z^k$ 
must equal to zero for $z=z_0$, this yields the equality $\frac{\rho}{r}(z_0-1)=1-z_0^{-k}$ and therefore equation {\ref{g}} becomes
\begin{equation} \label{h}
P_n=\frac{\rho}{rk}(z_0^{k-n-1})(z_0-1)
\end{equation}
The equation {\ref{h}} is the required expression for probability distribution.\\

\textbf{\emph{Validation:}} We can validate our derived expression by reducing it in the form of standard queueing model.
The CliqueCloak system resembles 
M/M/1 bulk service system, with a deviation that there is the additional concept of success anonymization.  That is there is an additional supplementary
parameter $r$ which is the probability of transition from one stage to another in case of success, otherwise it will remain in the same stage.
At this point we are able to derive the performance metrics for CliqueCloak approach.Therefore if the value of $r$ is 1, i.e the probability 
of success is 100 percent then the system will reduce to standard M/M/1 bulk service system. Substituting $r=1$, $\rho=\lambda/\mu$ and
$\frac{\lambda}{\mu}(z_0-1)=1-z_0^{-k}$ in equation {\ref{h}} we obtain the following expression.
\begin{equation} 
P_n=\frac{z_0^{k-n-1}}{k}(1-z_0^{-k})
\end{equation}
Notice that the above equation is the expression for M/M/1 queueing model with bulk service. Thus we validate our derivation for 
CliqueCloak system.

\subsubsection{\bf {Derivation of performance matrices}}
Using the probability distribution computed in the previous section, we can derive the required performance matrices.  
\begin{itemize}
\item {\bf Average Request in the system : } It is the average number of customer in the system including the request being processed by the anonymizing server denoted by $L$. It can be calculated using the derived expression of $P_n$ as  $L=\sum_{n=0}^{\infty}nP_n$. The expression for average number of customer in the system is given below.
       
\begin{equation}
 L=\frac{\lambda}{{\mu}rk}[k-\frac{1}{(z_0-1)}]
\end{equation}
Once we obtained the expression for average number of customer in the system $L$, we can derive the remaining performance matrix using $L$.
 \item {\bf Mean Queue Length:} The average number of requests in queue will be given as  $L_q=L-\rho$,     
\begin{equation}
 L_q=\frac{\lambda}{{\mu}rk}[(k-1)-\frac{1}{(z_0-1)}]
\label{eq-clickcloak-queue}
\end{equation}

\item {\bf Mean Sojourn Time:} The average time a request spends in the system called the sojourn time is denoted by $W$. The sojourn time is equal to the mean waiting time plus mean processing time for the request, i.e.       
\begin{equation}
 W=\frac{1}{{\mu}rk}[k-\frac{1}{(z_0-1)}]
\end{equation}

\item {\bf Mean Waiting Time:} The average time a request waits in queue for being processed in by anonymizer will be $W_q=W-1/\mu$,       
\begin{equation}
 W_q=\frac{1}{{\mu}rk}[(k-1)-\frac{1}{(z_0-1)}]
\label{eq-clickcloak-wait}
\end{equation}
\item {\bf Server Utilization:} It is the ratio of the rate at which request enters the system to the rate at which the
system can process the request.      
\begin{equation}
 S=\frac{\lambda}{{\mu}rk}
\end{equation}
\end{itemize}

\section{Experimental Results}\label{expts}
In this section we firstly discuss the experiments carried out to validate the model by comparing experimental results to their respective theoretical 
values obtained from the model. Further we evaluate the privacy approaches that we have modeled in terms of the queuing theory performance metrics -
 queue length and average turnaround time or sojourn time. In the experiments, requests are generated according to exponential distribution with parameter $\lambda$ set to 5, 
 and the service rate $\mu$ value fixed at 10. 
We vary the value of $\lambda$ but it does not exceed the value 10. The parameters used in our experiments is summarized in
Table {\ref{t}}. 

\begin{table}[h!] 
\caption{Experimental parameters with default values }
\label{t}
\centering
\begin{tabular}{{|c|l|}}
\hline
\textbf{Parameters} & \textbf{Default Values} \\ 

  \hline
\hline
 $\lambda$ & 5 queries per second\\ \hline
 $\mu$ & 10 queries per second\\ \hline
 k & 3 \\ \hline
  $z_0$ & 2.9196\\ \hline
  r & .33\\[1ex]
  \hline
\end{tabular}
\end{table}
 \subsection{Experimental validation of model}
\begin{figure}[h]
\begin{center}
\includegraphics[scale=0.21]{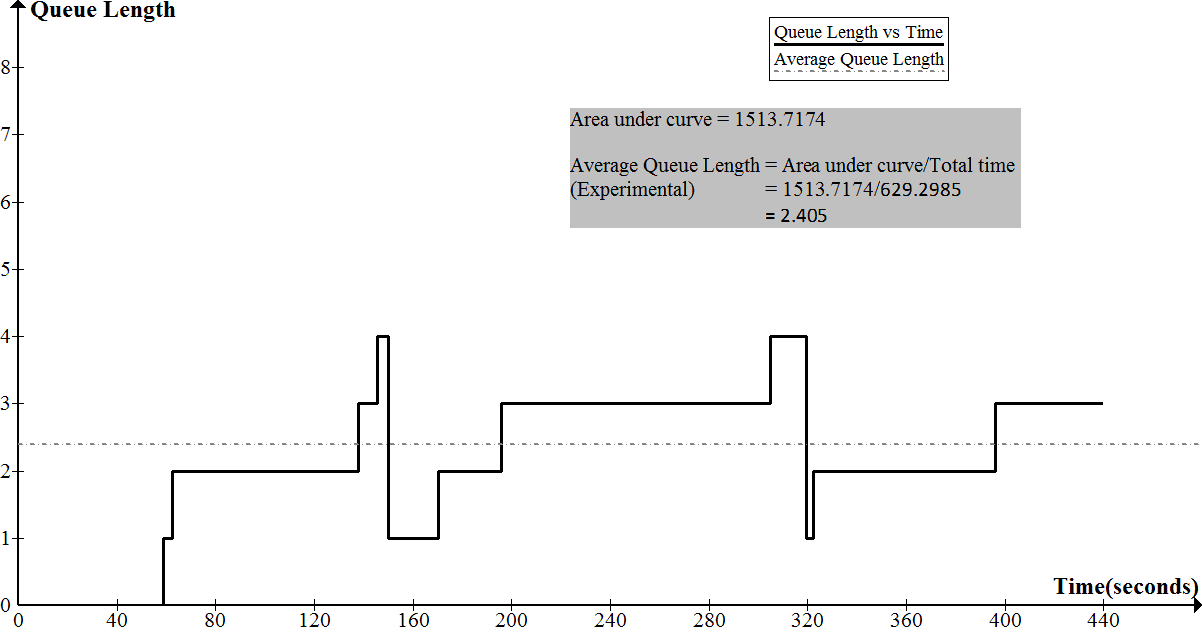}	
 \caption {ClickCloak: Change in queue length  with time}
 \label{instant-queue-click}
 \end {center}
\end {figure}

In order to compute the experimental average queue length, we record how the queue 
length varies with each arrival and departure as shown by solid line in Figure \ref{instant-queue-click}. The queue length increases by 1 for each arrival and decreases by $k$ for each departure as 
ClickCloak follows a bulk processing model of size $k$. In this experiment we took the value of $k$ to be 3. The processing will start only after three queries are in 
the queue. In the figure, the first query arrives at time 39 secs, the second query at time 41 secs and the third query at 
time 139 secs. However, the three arrivals do not form a clique. The fourth arrival occurs at time 143 secs and at time 
145 secs three queries leave the system.In this way curve for instantaneous queue length is constructed. Now the average queue length can be calculated using the curve as shown below:
 
\begin{equation}
 L_{Experimental}=\frac{Area\hspace{.05in} Under\hspace{.05in} Curve}{Total \hspace{.05in}Time}
\end{equation}
The dotted line in  Figure \ref{instant-queue-click} denotes experimental average queue length for default parameter values.
The computation of Experimental sojourn time is quite apparent.Experiment for n queries is conducted to record their sojourn times and mean is calculated. 
 The theoretical queue length and sojourn time for Casper is computed using Equations \ref{eq-clickcloak-queue} and \ref{eq-clickcloak-wait}.
respectively.

\begin{figure}[h!]
\begin{center}
\includegraphics[scale=0.7]{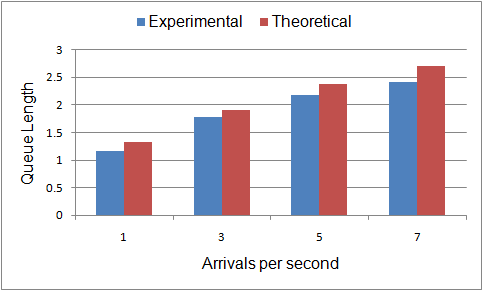}	
 \caption {ClickCloak: Comparison of queue length from theoretical model and experimental results}
  \label{click-queue}
 \end {center}
\end {figure}

\begin{figure}[h!]
\begin{center}
\includegraphics[scale=0.7]{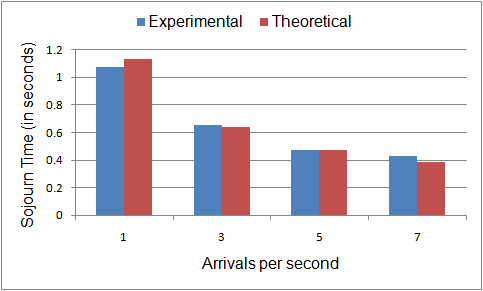}	
 \caption {ClickCloak: Comparison of sojourn time from theoretical model and experimental results}
  \label{click-wait}
 \end {center}
\end {figure}

The comparison of queue length and sojourn time from theoretical model as well as experimental results for different values of $\lambda$ is shown in Figure \ref{click-queue} and \ref{click-wait} respectively. The value of $\lambda$ is kept below 10 since our service rate $\mu$, which is system dependent, is fixed at 10 
and for a queueing model to be stable, arrival rate must be less than service rate. 
 The graph shows that the 
theoretical model closely follows the experimental results. The average error in result in case of queue length is about 
1.3 percent and for sojourn time the average error is 0.3 percent.

\subsection{Graphical Analysis of model}
\begin{figure}[h!]
\begin{center}
\includegraphics[scale=0.2]{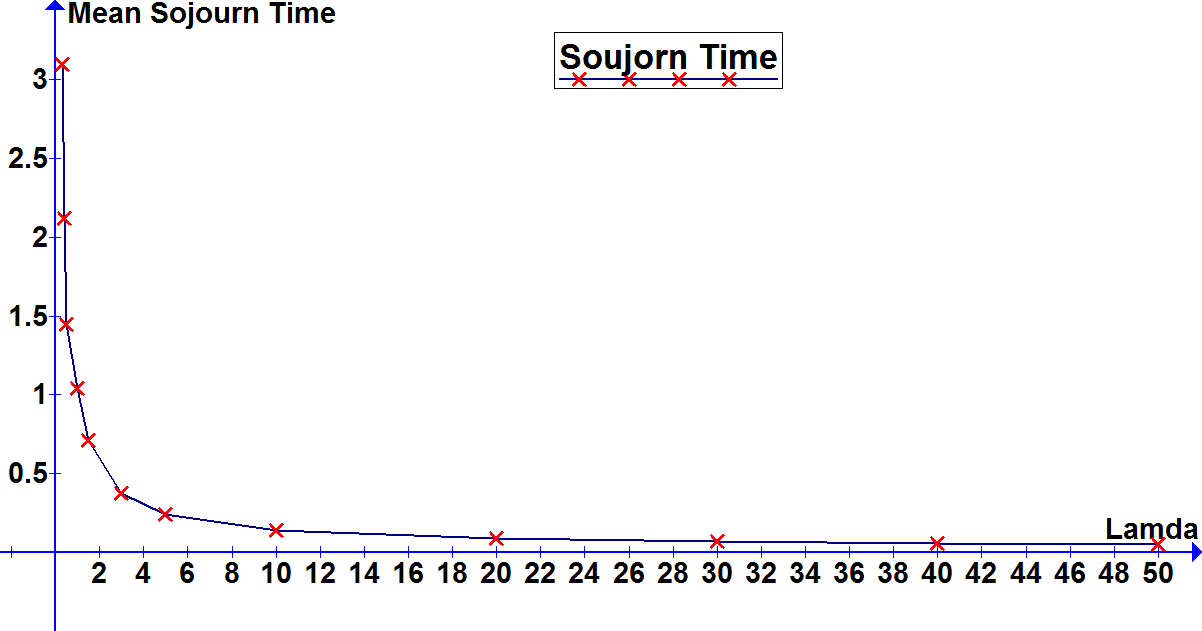}	
 \caption {ClickCloak: Characteristics graph of Sojourn Time}
  \label{click-wait2}
 \end {center}
\end {figure}

\begin{figure}[h!]
\begin{center}
\includegraphics[scale=0.2]{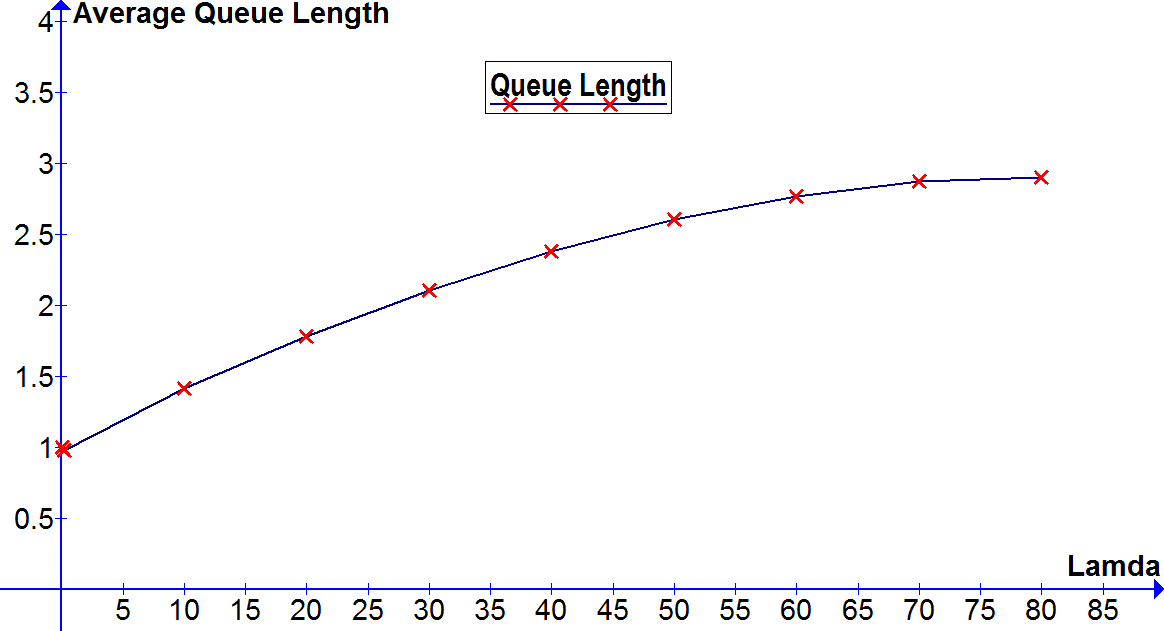}	
 \caption {ClickCloak: Characteristics graph of Queue Length}
  \label{click-queue2}
 \end {center}
\end {figure}

\begin{figure*}[h]
\begin{center}
\includegraphics[scale=0.67]{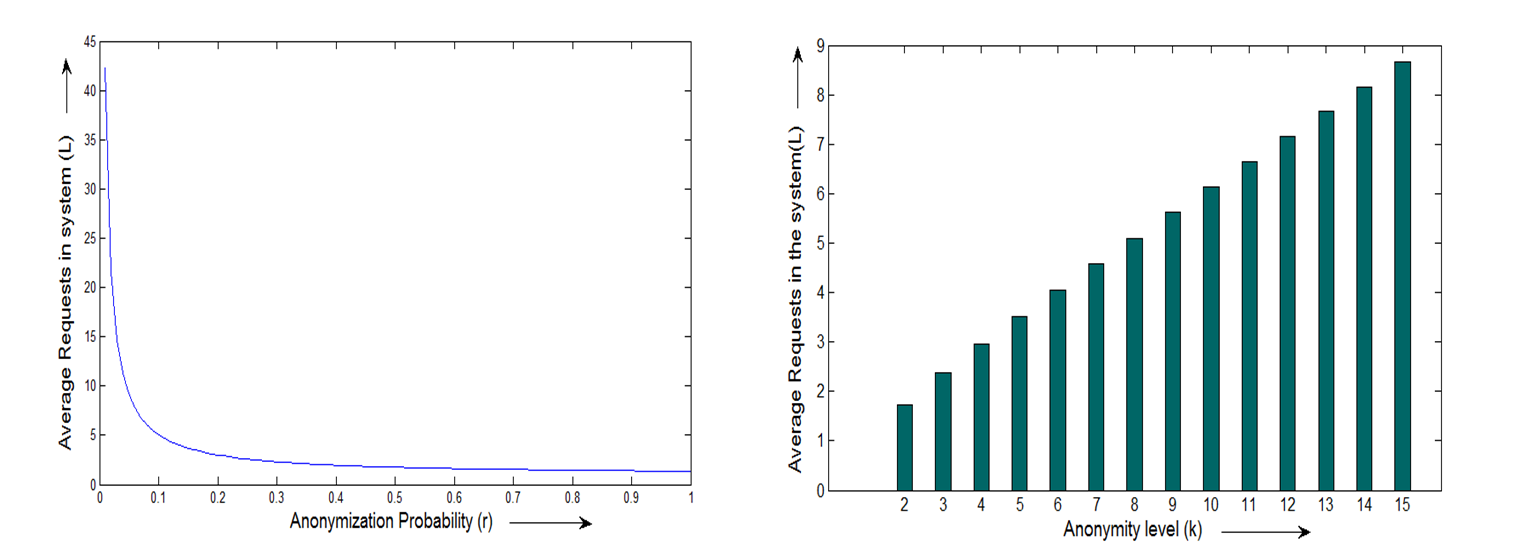}	
 \caption {Average queue length for different values of (a)Anonymization Probability (r) and (b)Anonymity level (k)}
 \label{1}
 \end {center}
\end {figure*}

\begin{figure*}[h]
\begin{center}
\includegraphics[scale=0.67]{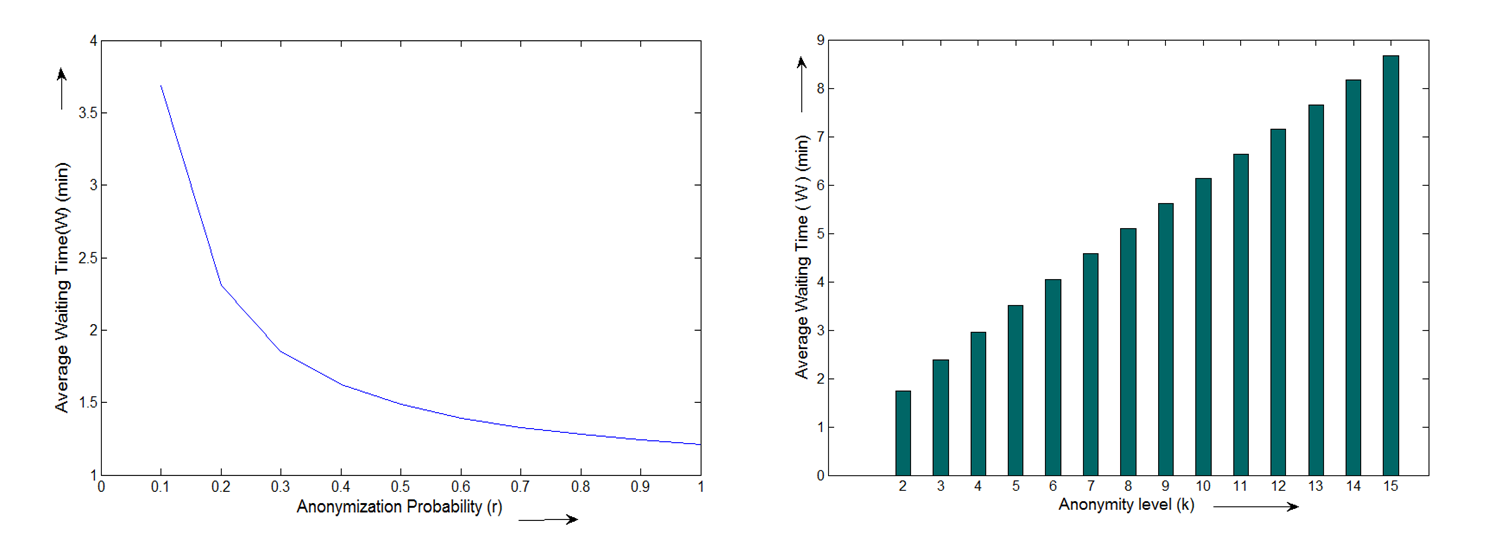}	
 \caption {Average Waiting time for different values of (a)Anonymization Probability (r) and (b)Anonymity level (k)}
 \label{2}
 \end {center}
\end {figure*}

\begin{figure*}[h]
\begin{center}
\includegraphics[scale=0.67]{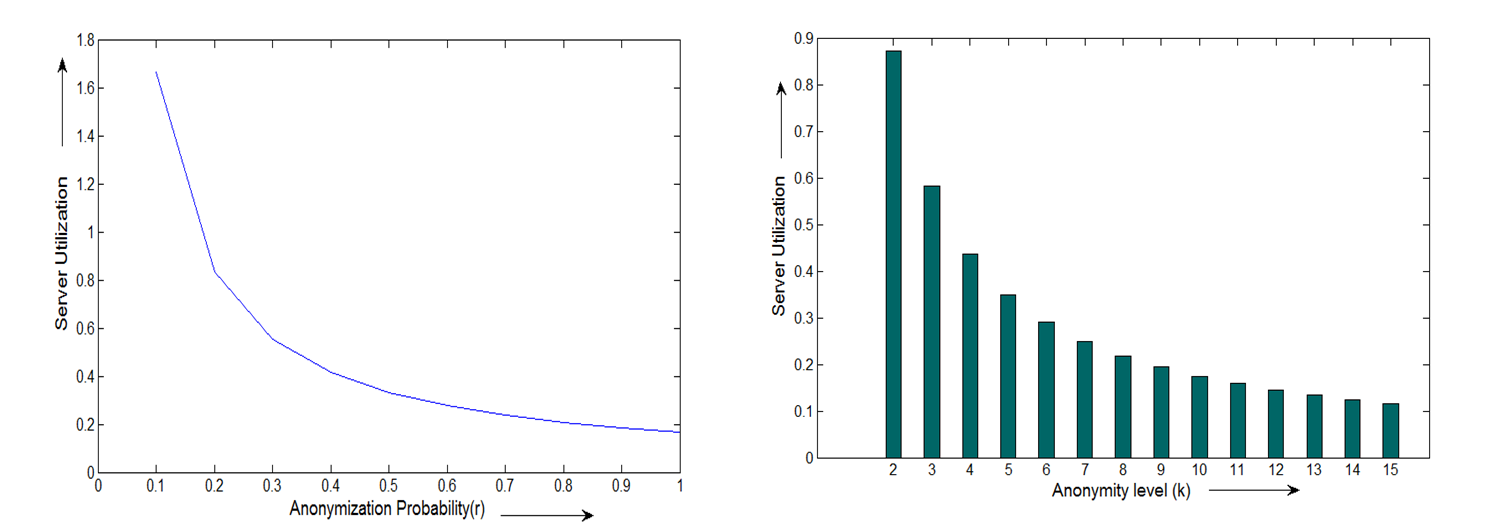}	
 \caption {Server Utilization for different values of (a)Anonymization Probability (r) and (b)Anonymity level (k)}
 \label{3}
 \end {center}
\end {figure*}

To analysis the characteristics of the system, graphs are constructed based on model.
Figure \ref{1}, \ref{2} and \ref{3} describes the effect of anonymization probability $r$ and privacy level $k$ on 
average queue length, average sojourn time and Server Utilization respectively.With increase in $k$, the cloaking time or processing time
increases. Consequently the average queue length and average sojourn time increases but the server utilization decreases as the server remain 
idle until server have the required number of $k$ queries. 
As can be seen from the Figure \ref{click-queue}, the queue length increases with arrival rate. The queue 
length for a arrival per second is a little more than 1, where as for 7 arrivals per second the queue length is about 2.5. 
This is  because queue length is directly proportional to $\lambda$, as given in Equation  \ref{eq-clickcloak-queue}. 
  On the other hand, we find that the sojourn time decreases with arrival as shown in Figure \ref{click-wait}. The sojourn time is about 1.1 seconds for 1 arrival 
per second and it reduces to 0.4 seconds for 7 arrivals per second. The reason behind this is that with increase in arrival rate, 
the anonymizing probability $r$ increases and  sojourn time being inversely proportional to $r$ (Equation  \ref{eq-clickcloak-wait}) 
is reduced. These results are further magnified and shown in the characteristics graph in Figure \ref{click-wait2} which shows 
how the sojourn time exponentially  reduces with linear increase in $\lambda$. However, after the arrival rate crosses a threshold, we find that the sojourn time becomes constant as it solely depends on the processing rate. The queue length on the other hand increases 
with increase in value of $\lambda$ as shown in Figure \ref{click-queue2}, the trend in increase is not exponential but can be 
approximated by a moving average. 


\subsection{Summary of Results} The following significant observations are noticed from our proposed queueing model as well as experimental results. 
\begin{enumerate}

\item{Unlike the top-down and bottom-up privacy approach whose processing time depends on user density, the query processing time for the bulk processing  approaches depend on the query arrival rate. With increase in arrival rate, the average query processing time
as well as average sojourn time is reduced. The query processing time for the bulk processing ClickCloak approaches depend on the arrival rate. With increase in arrival rate, the average query processing time as well as average sojourn time is reduced. }  

\item{Although the ClickCloak algorithms have the nice property of generating cloaking area with higher spatial and temporal resolutions and processing queries in bulk, they have a higher turn around time. The request-to-response time further increases with decrease in arrival rate. Thus the other additional objectives should be to minimize the queue length and turn around time.}
\end{enumerate}

\section{Conclusion and Future Work}\label{conclusion}
In this work, we propose a queueing theory based model to analyze the performance of privacy approaches of location based services. We used the model to analyze some well known graph based \textit{k}-anonymity  privacy approaches. The basic task involved in modeling these privacy approaches involves modeling the processing time or service rate. We show that these privacy approaches can be modeled using a variant of the bulk service queueing model. Service rate of the bulk processing class of ClickCloak
algorithms depend on the arrival rate of queries and an anonymizing probability \textit{r}. Due to the presence of \textit{r}, the ClickCloak algorithms cannot be directly modeled by the standard bulk processing. We derived a queueing model for these class of algorithms and validated the model by showing that it reduces to the standard bulk processing model when \textit{r} is set to 1. Experimental results also show the correctness of
all our proposed model. Our queueing theory based analysis of privacy approaches give meaningful insight into these algorithms which were previously not known.

Our study continues in various directions. One such study is to examine the behavior, if the systems do not follow the queueing model. A second interesting study would be to use the proposed framework to study the request-to-response time of real LBS applications under different user conditions.


\begin{thebibliography}{1}

 \bibitem{survey1}
M.F.~Mokbel , \emph{Privacy in Location-based Services: State-of-the-art and Research Directions},
\hskip 1em plus  0.5em minus 0.4em\relax  IEEE International Conference on Mobile Data Management, 2007.

 \bibitem{survey2}
 Yelp(2016)
 \textit{"http://www.yelp.com/"}
Online; accessed 15-Aug-2016
 

\bibitem{C1}
G.~Myles, A.~Friday and N.~Davies, \emph{Preserving Privacy in Environments with Location-based Applications}, 
\hskip 1em plus  0.5em minus 0.4em\relax IEEE Pervasive Computing, vol.2,no.1,pp 56-64, 2003.

\bibitem{C2}
M.~Youssef, V.~Atluri and N. R.~Adam, \emph{Preserving Mobile Customer Privacy: An Access Control System for Moving Objects and Custom Profiles'}, 
\hskip 1em plus  0.5em minus 0.4em\relax 6th Intl. Conf. on Mobile Data Management (MDM), 2005.

\bibitem{C3}
B.~Gedik and L.~Liu, \emph{Location Privacy in Mobile Systems: A Personalized Anonymization Model}, 
\hskip 1em plus  0.5em minus 0.4em\relax In Proceedings of the 25th IEEE international conference on distributed computing systems (ICDCS), pp 620–629,2005.

\bibitem{CliqueCloak}
B.~Gedik and L.~Liu, \emph{Protecting location privacy with personalized k-anonymity: architecture and algorithms},
\hskip 1em plus  0.5em minus 0.4em\relax IEEE Transactions on Mobile Computing, vol.7,no.1,pp 1-18,2008.

\bibitem{iClique}
X.~Pan and J.~Xu, \emph{Protecting Location Privacy against Location-Dependent Attacks in Mobile Services},
\hskip 1em plus  0.5em minus 0.4em\relax IEEE Transactions on Knowledge and Data Engineering, vol 24, no. 8, pp 1506-1519, 2008.

\bibitem{book}
L.~Kleinrock, \emph{Queueing System: Theory},
\hskip 1em plus  0.5em minus 0.4em\relax  Wiley, vol.1,pp 1-417,1975.

\bibitem{Casper}
M.~MF, C.~CY and A.~WG, \emph{The new casper: query processing for location services without compromising privacy},
\hskip 1em plus  0.5em minus 0.4em\relax In Proceedings of the 32nd International Conference on Very Large Databases (VLDB), Seoul, Korea, pp 763-774,2006.

\bibitem{bottom-up2}
P.Y. ~Li , W.C.~Peng , T.W.~Wang , W.S.~Ku , J.X.~Xu , J.A.~Hamilton, \emph{A Cloaking Algorithm based on Spatial Networks for Location Privacy},
\hskip 1em plus  0.5em minus 0.4em\relax In Proceedings of IEEE International Conference on Sensor Networks, Ubiquitous, and Trustworthy Computing, pp. 90-97, 2008.

\bibitem{Quadivison}
M.~Gruteser and D.~Grunwald, \emph{Anonymous usage of location based services through spatial and temporal cloaking},
\hskip 1em plus  0.5em minus 0.4em\relax  In Proceedings of the 1st international conference on mobile systems, applications and services (MobiSys),San Francisco, California, pp 31-42, 2003.

\bibitem{top-down2}
Y.~Ayong, Y.~Li and L.~Xu, \emph{A novel location privacy-preserving scheme based on l-queries for continuous LBS},
\hskip 1em plus  0.5em minus 0.4em\relax  Computer Communications, 2016.


\bibitem{PrivacyGrid}
B.~Bamba, L.~Liu, P.~Pesti and T.~Wang, \emph{Supporting Anonymous Location Queries in Mobile Environments with Privacygrid}, 
\hskip 1em plus  0.5em minus 0.4em\relax In Proceedings of the 17th international conference on World Wide Web (WWW '08). ACM, New York, NY, USA, 237-246, 2008.

 \bibitem{Historical}
C.~Bettini, S.~Mascetti, X.S.~Wang, D.~Freni and S.~Jajodia, \emph{Anonymity and Historical-Anonymity in Location-Based Services. In Privacy in Location-Based Applications},
\hskip 1em plus  0.5em minus 0.4em\relax Lecture Notes In Computer Science, Vol. 5599. Springer-Verlag, Berlin, Heidelberg 1-30, 2009.

\bibitem{Historical2}
A.I.~Saleh,A.~Ali-Eldin and A.A.~Mohamed, \emph{Historical based location management strategies for PCS networks},
\hskip 1em plus  0.5em minus 0.4em\relax Wireless Networks, Springer, 1-26, 2016.

\bibitem{Hilbert}
P.~Kalnis, G.~Ghinita, K.~Mouratidis and D.~Papadias,  \emph{Preventing Location-based Identity Inference in Anonymous Spatial Queries},
\hskip 1em plus  0.5em minus 0.4em\relax  IEEE Transactions on Knowledge and Data Engineering 19(12), 1719–1733, 2007.

\bibitem{Hilbert2}
N.~Cui, X.~Yang  and B.~Wang, \emph{A Novel Spatial Cloaking Scheme Using Hierarchical Hilbert Curve for Location-Based Services},
\hskip 1em plus  0.5em minus 0.4em\relax Web-Age Information Management: 17th International Conference, WAIM 2016, Springer International Publishing, 15-27, 2016.

\bibitem{map-aware}
M.L.~Damiani, E.~Bertino and C.~Silvestri, \emph{Protecting Location Privacy against Spatial Inferences: the PROBE Approach},
\hskip 1em plus  0.5em minus 0.4em\relax  In Proceedings of the 2nd SIGSPATIAL ACM GIS 2009 International Workshop on Security and Privacy in GIS and LBS (SPRINGL '09). ACM, New York, NY, USA, 32-41, 2009.


\end{thebibliography}

\FloatBarrier 
\balance

\end{document}